\documentclass[%
 aip,
 sd,%
 amsmath,amssymb,
 reprint,%
]{revtex4-1}

\usepackage{graphicx}
\usepackage{siunitx}
\usepackage{dcolumn}
\usepackage{bm}

\begin{document}

\preprint{AIP/123-QED}

\title[Phase noise cancellation in polarisation-maintaining fibre links]{Phase noise cancellation in polarisation-maintaining fibre links}

\author{B. Rauf}
 \email{b.rauf@inrim.it}
 \altaffiliation[Also at ]{Dipartimento di Elettronica e Telecomunicazioni, Politecnico di Torino, Corso duca degli Abruzzi 24, 10129 Torino, Italy}
\author{M. C. V{\'e}lez L{\'o}pez}%
\altaffiliation[Also at ]{Real Instituto y Observatorio de la Armada, Calle Cecilio Pujazon, 11100 San Fernando, Cadiz, Spain}
\author{P. Thoumany}
\author{M. Pizzocaro}
\author{D. Calonico}
\affiliation{Physical Metrology Division, Istituto Nazionale di Ricerca Metrologica (INRIM), Strada delle Cacce 91, 10135 Torino, Italy}%

\date{\today}

\begin{abstract}
The distribution of ultra-narrow linewidth laser radiation is an integral part of many challenging metrological applications. Changes in the optical pathlength induced by environmental disturbances compromise the stability and accuracy of optical fibre networks distributing the laser light and call for active phase noise cancellation. Here we present a laboratory scale optical (at $\SI{578}{nm}$) fibre network featuring all polarisation maintaining fibres in a setup with low optical powers available and tracking voltage-controlled oscillators implemented. The stability and accuracy of this system reach performance levels below $1\times10^{-19}$ after $\SI{10 000}{s}$ of averaging.
\end{abstract}

\pacs{42.81.-i}

\keywords{fibre links, phase noise stabilisation, polarisation-maintaining fibre}
                          
\maketitle

\section{\label{sec:Introduction}Introduction}

Optical fibres are an essential tool of modern metrology and spectroscopy experiments. They offer great flexibility and are able to deliver high mode-qualities and stable polarisation at their output, but also transfer environmental noise, such as seismic/acoustic vibrations or temperature fluctuations, into optical phase noise. This noise can be cancelled out by referencing the fibre output light to the incoming radiation and using this signal to correct the phase and frequency of the light at the output \cite{Ma1994}.

Optical fibre links offer the possibility to deliver light of superb stability and accuracy, even across continental distances \cite{Lisdat2016,  Calonico2014, Matveev2013}. This allows the transfer of ultra-stable optical radiation between remote locations, enabling radio-astronomy with unprecedented resolution \cite{Clivati2017} as well as sub-Hz spectroscopy in atomic clocks \cite{Takamoto2015}. Fibre links are also essential parts of high stability measurements of optical frequency ratios as a prerequisite to a redefinition of the second in the International System of Units (SI) \cite{Gill2016} and relativistic geodesy \cite{Lisdat2016, Takano2016}, as well as the transfer of spectral purity between ultra-stable clock lasers \cite{Akatsuka2014}, even across diverse wavelengths \cite{Falke2014}.

The most accurate frequency standards to date are optical atomic clocks, reaching uncertainties of a few parts in $10^{18}$, although they are not yet realising the definition of the SI-unit the second \cite{Gill2016}. Various candidates for a future redefinition are being investigated \cite{Ludlow2015}, while optical clocks are expanding their fields of use to quantum simulations \cite{Livi2016}, geodesy and relativity experiments as well as applications \cite{Bjerhammar1985, Lisdat2016, Takano2016} and  the search for variations of fundamental constants \cite{Blatt2008, Godun2014}.

In all optical atomic clocks an ultra-stable laser interrogates a sub-Hz linewidth optical transition. This laser is usually locked to an ultra-stable reference, such as an ultra-stable cavity. Great efforts are being undertaken to develop cavities reaching thermal noise floor limited stabilites below $1\times10^{-16}$ \cite{Matei2017, Haefner2015}. Such spectral purity is usually preserved throughout the optical path of the laser by phase noise cancelled optical fibre links. Therefore the requirement for these links is to have a stability better than $5\times10^{-17}\tau^{-1/2}$, as obtained in contemporary performances of optical cavities \cite{Matei2017} and optical lattice clocks \cite{Schioppo2016}, and an accuracy below the best clock uncertainty evaluations in the low $10^{-18}$ region \cite{Gill2016}.\\
 At the ``Istituto Nazionale di Ricerca Metrologica'' (INRIM) an $^{171}$Yb optical lattice clock is in operation. The ultra-narrow double-forbidden $^1S_0\rightarrow{}^3P_0$ transition is interrogated by a $\SI{578}{nm}$ laser stabilised to an ultra-stable cavity. A complete description of the experiment can be found in \cite{Pizzocaro2017}.\\
Here we report on the phase noise cancellation in a laboratory scale setup at the INRIM facilities distributing the visible ultra-stable laser using polarisation-maintaining (PM) optical fibres. The use of PM fibres is common in laboratory experiments that demand a stable polarisation state of the light, such as high precision atomic spectroscopy, and where power fluctuations arising from polarisation to amplitude noise conversion need to be avoided. With standard single mode fibres this is commonly achieved by using manual polarisation controls that require to be tweaked as necessary. Yet such a configuration can still be quite susceptible to environmental changes.
In our setup the optical power circulating inside the fibres can be as low as tens of micro watts. Therefore tracking voltage-controlled oscillators (VCO) are engaged in the generation of the phase noise error signal, enhancing the robustness of the system. We show that our system meets the demands of state-of-the-art optical clocks and reaches a performance similar to the established single mode fibre links \cite{Grosche2014}.

\section{\label{sec:Setup}Setup}

\begin{figure}
\includegraphics[scale=0.235]{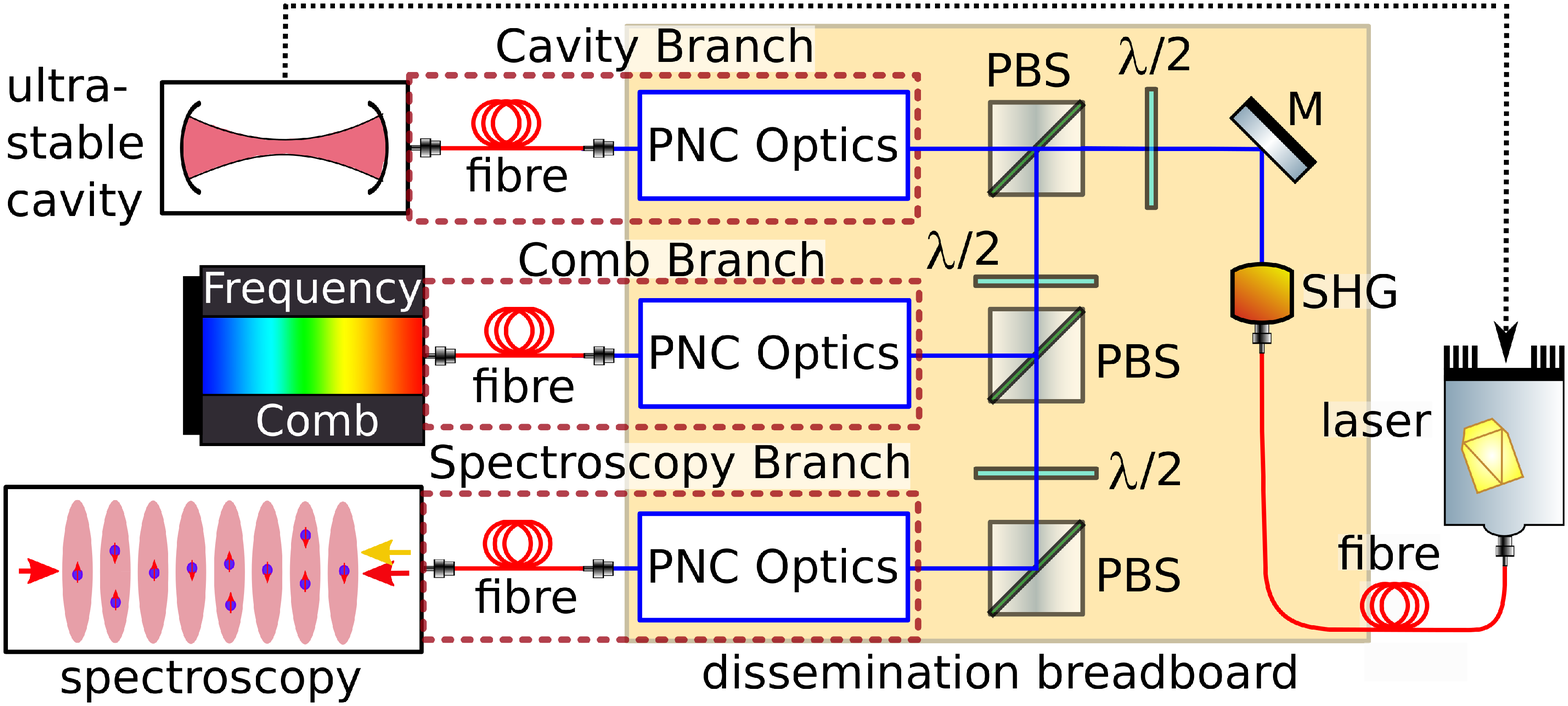}
\caption{\label{fig:Schemegen} Dissemination of the $\SI{578}{nm}$ clock laser. The laser light is distributed to the ultra-stable cavity, the frequency comb and the atomic spectroscopy by phase-noise cancelled (PNC) fibre links (called branches, inside red dotted boxes).  SHG = second harmonic generation,  M = mirror, PBS = polarising beam splitter, $\lambda/2$ = half-wavelength retardation plate, blue lines follow the free-space optical path. The black dotted line denotes DC signals of the laser stabilisation to the ultra-stable cavity.}
\end{figure}

A scheme of the $\SI{578}{nm}$ dissemination is shown in fig. \ref{fig:Schemegen}. The laser itself is situated in a neighbouring laboratory to the Yb lattice clock and is a frequency-doubled $\SI{1156}{nm}$ diode laser. The total optical power available at $\SI{578}{nm}$ after the doubling-crystal is about $\SI{7}{mW}$. A large part of this power (80\%) is sent towards the frequency comb in order to reach a high signal-to-noise ratio in the beatnote with the comb light. The spectroscopy of the ultra-narrow atomic transition and the ultra-stable cavity on the other hand need only little power, as the light power impinging on the cavity and the Yb atoms is in the $\SI{}{\micro\watt}$ to $\SI{}{\nano\watt}$ region.  Therefore the fibre noise cancellation system needs to work even with low optical powers (about $\SI{100}{\micro\watt}$) circulating inside the optical path of the phase noise cancellation system, that is the power in each branch after the polarising beam splitter (PBS) in fig. \ref{fig:Schemegen}. A fraction of this power is then available for the generation of the optical phase noise error signal. The laser light is connected to the dissemination breadboard by optical fibre. Any airflow inside the lab is shielded by an acrylic enclosure around the breadboard protecting the free-space optical paths of the setup. 

The fast stabilisation of the diode laser to one cavity mode acts upon the laser current and is supported by a low-bandwidth drift compensation controlling the diode laser internal cavity length (piezo-voltage). The performance of the laser stabilised to the ultra-stable cavity is similar to what has been previously reported \cite{Pizzocaro2012}. After passing through the second harmonic generation (SHG) crystal the radiation is split into three branches, each leading to a dedicated $\SI{20}{m}$ PM fibre delivering the light to the ultra-stable cavity, the frequency comb and the atomic spectroscopy within the science chamber of the lattice clock. 

These PM fibres are optimised for wavelengths around the Yb clock transition of $\SI{578}{nm}$ and engineered in the panda style, meaning that a significant fibre birefringence is obtained through stress rods pulling on the core and thereby suppressing cross-talk between the fibre polarisation axis. As a consequence small variations of the fibre birefringence exerted e.g. by mechanical forces do not lead to significant polarisation noise, unlike the case of single-mode fibres \cite{Noda1986}. This attribute allows polarisation cleaning after the fibre without introducing polarisation to amplitude noise conversion.

\begin{figure}
\includegraphics[scale=0.243]{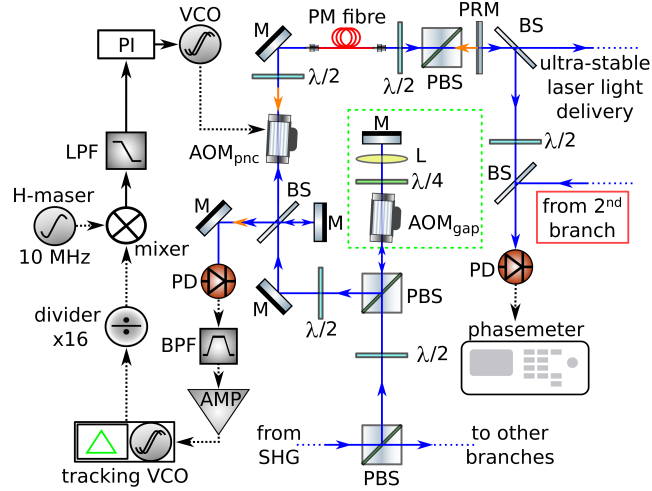}
\caption{\label{fig:schemebranchumeas} Sketch of the fibre noise cancellation setup and its assessment. The laser light enters the branch coming from the SHG crystal. The double-pass AOM in cat-eye configuration is marked with green dots. The laser light is signalled with blue arrows, but the retro-reflected beam path with orange arrows. PD = photodiode, VCO = voltage controlled oscillator, LPF = low-pass filter, AMP = amplifier, BPF = band-pass filter, L = lens, PRM = partial reflecting mirror, PI = proportional-integral controller, BS = power beam splitter}
\end{figure}
 
All of these branches share an identical layout, which is sketched in fig. \ref{fig:schemebranchumeas}.
 The ultra-stable laser light enters the branch coming from the direction of the SHG output, as shown in fig. \ref{fig:Schemegen}. At first a double-pass AOM ($AOM_{gap}$) in cat-eye configuration \cite{Donley2005} is used to bridge the frequency gap between the nearest cavity mode and the atomic resonance. 
 Subsequently a power beam splitter (BS) serves as the reference point of an imbalanced optical interferometer with the short arm constituting  the reference beam ($30\%$ of the incoming beam power reflected) for the generation of the phase-noise error signal. Before passing through the $\SI{20}{m}$ PM fibre (long interferometer arm) the beam is imprinted with an $\SI{80}{MHz}$ frequency shift by AOM\textsubscript{pnc}. A half-wave plate is used to align the light polarisation with the slow axis of the PM fibre, minimising polarisation-noise from the fibre. The polarisation extinction ratio at this point is determined by the PBS in the optical path after the cat-eye AOM setup as 100:1 (reflected beam).

After  the fibre the polarisation of the light is cleaned and a part ($\approx40\%$) of the radiation is retro-reflected by a partial reflecting mirror. Only the comb-branch features higher optical  power, compensated by a reduced reflectivity of the partial reflecting mirror ($\approx15\%$). Typically $\SI{20}{\micro\watt}$ of optical power are delivered after the partial reflecting mirror (in the case of the comb branch these are some $\SI{1.5}{\milli\watt}$). The polarisation is cleaned using a $\lambda/2$ waveplate and a PBS with a polarisation extinction ratio of at least $1000:1$. A maximised PBS output ensures that also the back-reflected beam is aligned on the slow axis of the PM fibre. This procedure suppresses amplitude noise in the optical interferometer stemming from the fibre birefringence, which is quite sensitive to temperature changes in the case of panda-style PM fibres \cite{Noda1986}.
 The reflected light passes once again the AOM\textsubscript{pnc} and then interferes with the reference beam.
 All our fibres use angle flat connectors (a variation of angle polished connectors manufactured by OZ optics) to reduce back-reflection from the fibre tip. We found that this back-reflection gives rise to a signal $\SI{30}{dB}$ smaller than the one from the retro-reflector.
 
We note that having the same polarisation on the retro-reflected beam significantly improves the system performances: in fact, we observed an enhanced sensitivity to polarisation fluctuations  when the back-reflected beam had a rotated polarisation direction with respect to the initial beam. This happens because this beam is then not aligned to a fibre axis and the birefringence is quite sensitive to temperature fluctuations.

The beatnote between the reference and the retro-reflected beam carries the information on the phase noise along the optical path and is found at a frequency of $\SI{160}{MHz}$. It is recorded on a fast photodiode (Menlo FPD510-FV, bandwidth $\SI{200}{MHz}$) and subsequently filtered with a band-pass filter (bandwidth $\SI{14}{MHz}$ around $\SI{157}{MHz}$) and then amplified by $\SI{30}{dB}$. This beatnote signal is then tracked by a voltage-controlled oscillator (VCO), which has a bandwidth of $\SI{3}{MHz}$. As a result the reliability of the fibre link is improved, since fluctuations in the optical beatnote might cause cycle slips otherwise. The VCO frequency is divided by 16 with a frequency divider and directly compared in a mixer with a $\SI{10}{MHz}$ reference signal coming from a H-maser. A low-pass filter (bandwidth $\SI{500}{kHz}$) subsequently filters out the DC part of this signal. A proportional-integral (PI) controller takes this error signal to feed the frequency of a VCO that in turn drives the AOM\textsubscript{pnc} in order to cancel all phase noise occurring along the optical and in-fibre path between the BS and the partial reflecting mirror. In order to minimise the residual noise along the optical path the control bandwidth of the PI is set individually for each fibre noise cancellation to optimised values around $\SI{20}{kHz}$. 

The radiation delivered by all three branches has the same spectral attributes except for a possible adjustable frequency offset introduced by the double-pass AOM and the intrinsic optical path-length noise of each branch.  

The fibre link electronics feature also an ``open loop'' operation. In this mode the VCO providing the AOM\textsubscript{pnc} RF signal at $\SI{80}{MHz}$ is stabilised with a phase locked loop to the H-maser signal. This operational mode is useful for initial alignment or to keep light delivered through the fibre even when the back-reflected signal is missing.

At last the part of the beam not retro-reflected after the partial reflecting mirror is used to deliver the ultra-stable radiation to its destination (the ultra-stable cavity, the frequency comb or the atomic spectroscopy). 

The facilities of the Yb lattice clock at INRIM feature an air conditioning system that stabilises the temperature inside the clock laboratory to within a periodic oscillation of $\SI{0.2}{K}$ amplitude and with a periodicity of $\SI{15}{min}$ (quasi sinusoidal oscillation), while the airflow alternates its intensity accordingly. This means that the momentary temperature change in the lab can be as high as $\SI{0.5}{\milli\kelvin\per\second}$. Such temperature oscillations  introduce changes in the fibre length and the fibre index of refraction leading to a calculated noise at the half-period time (maximum temperature excursion and applying the sensitivities of pure silica \cite{Toyoda1983}) of about $1\times10^{-17}$ and $1\times10^{-16}$, respectively.

There is a remaining up to  $\SI{2}{m}$ of free-space optical path in each of the Yb laser branches that is not shared by any other branch and is situated outside of the phase-noise cancellation interferometer. An enclosure made from acrylic glass around the breadboard containing the optics of these branches shields efficiently acoustic noise and short-term density fluctuations, whereas the stiff breadboard ensures that seismic noise is common to all optical elements. The dissemination breadboard, which is made from steel, experiences also temperature excursions due to the lab temperature fluctuations. These have the same periodicity as the laboratory temperature, but with a much smaller amplitude of $\SI{4}{mK}$.
The Doppler shift caused by the steel breadboard temperature fluctuations altering the free-space optical path is estimated at $1\times10^{-18}$ at $\SI{450}{s}$.

\section{\label{sec:Results}Results}

The metrological characterisation of the phase noise cancellation was performed through pairwise evaluations of the fibre links, as shown in fig. \ref{fig:schemebranchumeas}. For these measurements the radiation after the partial reflecting mirror was beat against the light arriving from a $2$\textsuperscript{nd} identical setup in a short free-space optical path. These measurements were performed inside a separate acrylic enclosure and the beatnote on a photodiode recorded with a phasemeter or frequency counter.

In a first step we determined the minimum input light power within one branch (after the PBS in fig. \ref{fig:Schemegen}) needed for the phase noise cancellation to function properly to be $\SI{90}{\micro\watt}$. Below this value cycle slips appeared in the locking of the tracking oscillator, compromising the stability of the phase noise cancellation.

\begin{figure}
\centering
\includegraphics[scale=0.32]{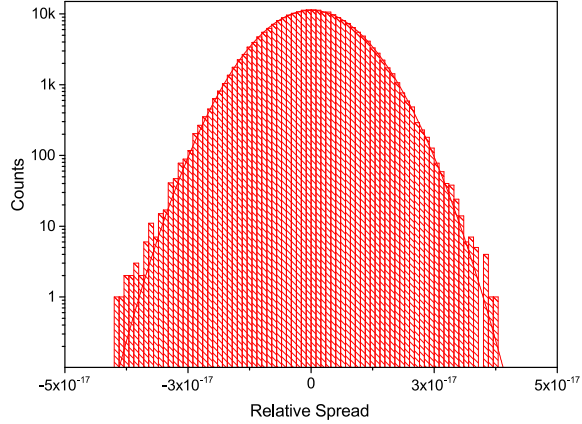}
\caption{\label{fig:Histogram}Histogram of a 20 hours measurement looking for cycle slips in the beatnote. No points were recorded outside of the scale used for the abscissa.}
\end{figure}

The possibility of such cycle slips was investigated by looking for jumps by an integer number of $2\pi$ in phase in the beatnote between the outputs of the phase noise cancelled fibres. As can be seen from fig. \ref{fig:Histogram}, no cycleslips occurred in 20 hours of measurement. All data points were within 0.05 Hz ($1$ part in $10^{16}$).

\begin{figure}
\centering
\includegraphics[scale=0.37]{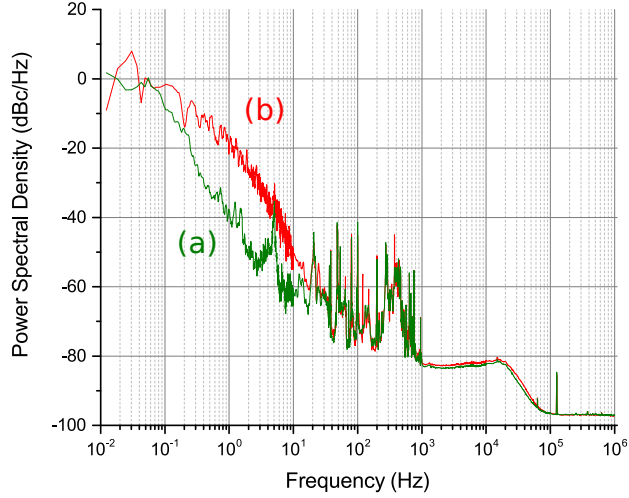}
\caption{\label{fig:Airflowpsd}Power spectral density (PSD) of the beatnote between two branches with (a) and without (b) the lid of the acrylic enclosure attached. The slightly higher phase noise level compared to fig. \ref{fig:PSDc} is due to different optical and electronic power levels during this  measurement. The results are nonetheless transferable to the current setup.}
\end{figure}

We compared the noise between two branches with and without the lid of the acrylic enclosure attached in order to determine the influence of turbulent air in the free-space region of the branches on the performance of the phase noise cancellation. The results are shown in fig. \ref{fig:Airflowpsd}. Turbulent air in the free-space paths outside of the phase-noise cancellation loops does increase the phase noise of the disseminated laser light in a region between $\SI{0.1}{Hz}$ and  $\SI{10}{Hz}$, which includes especially the timescale of one clock cycle ($\SI{0.1}{s}$ to $\SI{0.5}{s}$) and is therefore most critical. The acrylic enclosure suppresses this effect efficiently.

\begin{figure}
\centering
\includegraphics[scale=0.37]{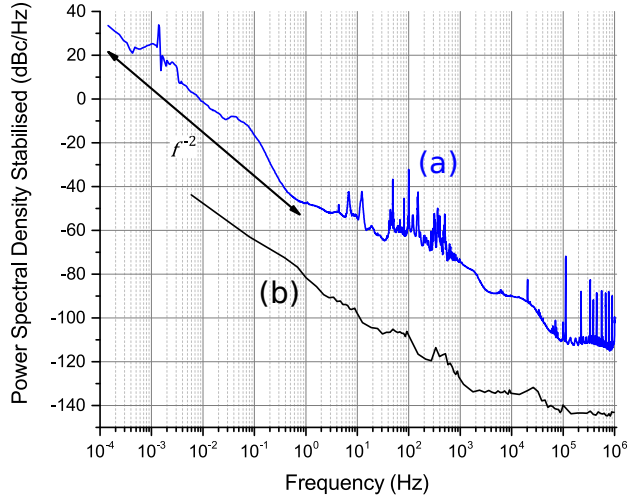}
\caption{\label{fig:PSDc}Power spectral density of the beatnote between the output laser radiation of two phase-noise cancelled fibre links, measured with a Microsemi 5125A phasemeter (a) and noise floor of the phasemeter (b)}
\end{figure}

The power spectral density of a typical phase noise cancellation performance measurement between two branches is shown in fig. \ref{fig:PSDc}. The curve follows a $1/f^2$ behaviour (white frequency noise) in the low frequency region ($\SI{0.1}{mHz}-\SI{0.1}{Hz}$), succeeded by - typically acoustic - noise displaying pronounced peaks around $\SI{10}{Hz}$ and between $\SI{300}{Hz}$ and $\SI{500}{Hz}$. Electronic noise stemming from the power line at $\SI{50}{Hz}$ and harmonics thereof is also visible. A white phase noise pedestal is reached for frequencies between $\SI{4}{kHz}$ and $\SI{20}{kHz}$. The SNR of the beatnote on the detection photodiode (\SI{42}{dB} in $\SI{100}{kHz}$) agrees with the measured phase noise ($\approx \SI{-90}{dBc\per\hertz}$) in this region. At $\SI{20}{kHz}$ the bandwidth of the control is reached and for higher frequencies only electronic noise is left in the signal.

\begin{figure}
\centering
\includegraphics[scale=0.39]{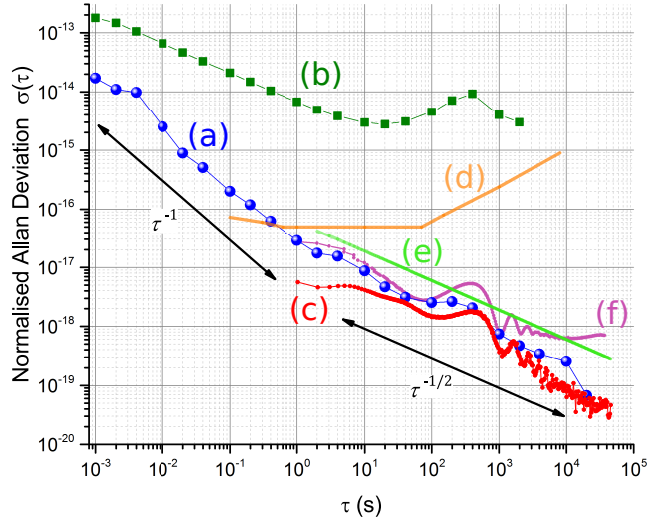}
\caption{\label{fig:Allanc}Allan-deviation of the same measurement as in fig. 5 (a), compared with measurements of the unstabilised link performance (b) and a $4$ days long assessment of the fibre link performance (c). The best reported stabilities of an optical cavity \cite{Matei2017} (d) and of an ensemble of optical lattice clocks \cite{Schioppo2016} (e) are shown as benchmarks. The magenta curve (f) displays the total Allan-deviation obtained when using a mismatched polarisation for the back-reflected beam. Lines between measurement points are there to guide the eye.}
\end{figure}

The Allan-deviation of the phase noise cancellation performance is shown in fig. \ref{fig:Allanc}. These measurements were obtained using a phasemeter with a bandwidth of $\SI{500}{Hz}$ and a gate time of $\SI{1}{ms}$. The exception is the 4 days measurement that was acquired with a high-resolution K+K FXE counter ($\Lambda$-variance) and a gate time of $\SI{1}{s}$ (the resulting equivalent bandwidth is $\SI{0.5}{Hz}$).

The unstabilised fibre link shows an Allan-deviation that is not compatible with state-of-the-art ultra-stable lasers and clock stabilities. Once the fibre noise cancellation is active the low integration time regime ($\tau<\SI{1}{s}$) averages down as $1/\tau$, matching white phase noise, as expected for fibre links.  At about $\SI{4}{s}$ the slope changes towards an overall $\tau^{-1/2}$ trend, revealing white frequency noise. Looking at the 4 days measurement it becomes clear that the noise at timescales $\tau\geq\SI{450}{s}$ is dominated by the periodic noise of the air conditioning system inside the INRIM laboratories.

The $\Lambda$-variance of the high resolution counter  reduces its sensitivity to white phase noise in comparison to the phasemeter measurements \cite{Dawkins2007, Calosso2016}.  As a consequence the Allan deviation obtained with the counter measurement is a bit below the phasemeter values for   those short timescales where white phase noise dominates. For integration times above a few seconds the noise is changing to white frequency noise behaviour and at long timescales the blue and the red curve agree quite closely, as expected. When taking a phasemeter measurement with a bandwidth of $\SI{0.5}{Hz}$, we saw that the blue and red curve do also agree at the $\SI{1}{s}$ value, but the information on the high frequency behaviour of the fibre link is lost. 

In conclusion we measured a stability of $2\times10^{-17}$ at $\tau=\SI{1}{s}$ that follows a $7\times10^{-18}\tau^{-1/2}$ behaviour for longer timescales ($\tau\geq\SI{1 000}{s}$), as shown in fig. \ref{fig:Allanc}. The phase noise cancelled fibre link has a higher stability than the benchmark curves representing the latest performances of ultra-stable lasers \cite{Matei2017} and optical clock ensembles \cite{Schioppo2016}.
The resulting uncertainty of $5\times10^{-20}$ is limited only by statistics. No frequency offset in the beatnote between the pairwise investigation of the fibre links was detected. 

The fibre noise cancellation system presented here maintains the same polarisation throughout the reference- and the back-reflected beam. If the polarisation of the back-reflected beam was intentionally not aligned with the fibre axis by approximatively $45°$, we observed a noise floor in the Allan-deviation at $6\times10^{-19}$ (magenta curve).

\section{\label{sec:Discussion}Discussion}

We have successfully developed and characterised a fibre noise cancellation system designed to tackle the need of spectral purity transfer of an ultra-stable laser at $\SI{578}{nm}$ using circulating optical powers below $\SI{100}{\micro\watt}$. The assessed fibre link uncertainty of $5\times10^{-20}$ and stability of $7\times10^{-18}\tau^{-1/2}$ surpass the best contemporary lasers stabilised to an ultra-stable cavity \cite{Matei2017}, as well as the highest demonstrated stability and accuracy of an optical clock so far \cite{Schioppo2016, Gill2016}.

The periodic change of temperature acting on the length of the optical path between the fibre noise cancellation units leads to an uncertainty in the low  $10^{-18}$ region at $\SI{450}{s}$, in agreement with our predictions. The temperature change follows a somewhat sinusoidal curve and therefore, for timescales $\tau\gg\SI{450}{s}$, turns into a white frequency noise behaviour, averaging down as $\tau^{-1/2}$. 

The performance we measured is suitable for the transfer of high-purity laser radiation and for state of the art optical clocks. If desired it could still be  diminished by a reduction of the optical path between the dissemination branches (fig. \ref{fig:Schemegen}), to the point of using the same reference mirror for all fibre links (calling for a sophisticated electronic setup), or by thermally insulating the steel breadboard supporting the whole free-space optics of the laser dissemination and the fibre noise cancellation systems.

\begin{acknowledgments}
The authors would like to thank C. Clivati and E. Bertacco for their support and helpful discussions.
The authors acknowledge funding from the, from the Innovative Training Network (ITN) Future Atomic Clock Technology (FACT), and from the European Metrology Programme for Innovation and Research (EMPIR) project 15SIB03 OC18. This project has received funding from the EMPIR programme co-financed by the Participating States and from the European Union’s Horizon 2020 research and innovation programme.
\end{acknowledgments}

\bibliography{aipsamp}

\end{document}